# Worldwide Energy Harvesting Potential of Hybrid CPV/PV Technology

Juan F. Martínez[*], Marc Steiner, Maike Wiesenfarth, Henning Helmers, Gerald Siefer, Stefan W. Glunz and Frank Dimroth

Fraunhofer Institute for Solar Energy Systems ISE, Heidenhofstr. 2, 79110 Freiburg, Germany

[*]Author to check the proofs: Juan F. Martínez

Fraunhofer Institute for Solar Energy Systems, Heidenhofstr. 2, 79110 Freiburg, Germany

Email: juan.francisco.martinez.sanchez@ise.fraunhofer.de, Tel.: +49–761–4588-5072

# Abstract

Hybridization of multi-junction concentrator photovoltaics with single-junction flat plate solar cells (CPV/PV) can deliver the highest power output per module area of any PV technology. Conversion efficiencies up to 34.2% have been published under the AM1.5g spectrum at standard test conditions for the EyeCon module which combines Fresnel lenses and III-V four-junction solar cells with bifacial c-Si. We investigate here its energy yield and compare it to conventional CPV as well as flat plate PV. The advantage of the hybrid CPV/PV module is that it converts direct sunlight with the most advanced multi-junction cell technology, while accessing diffuse, lens-scattered and back side irradiance with a Si cell that also serves as the heat distributor for the concentrator cells. This article quantifies that hybrid bifacial CPV/PV modules are expected to generate a 25 - 35% higher energy yield with respect to their closest competitor in regions with a diffuse irradiance fraction around 50%. Additionally, the relative cost of electricity generated by hybrid CPV/PV technology was calculated worldwide under certain economic assumptions. Therefore, this article gives clear guidance towards establishing competitive business cases for the technology.

# 1. Introduction

This article investigates the worldwide potential of hybrid concentrator/flat plate photovoltaic technology (CPV/PV) in terms of annual electrical energy yield. The hybrid CPV/PV approach aims at complementing the highest conversion efficiency of direct sunlight, achieved by III-V concentrator multi-junction solar cells, with the additional absorption of diffuse and backplane irradiance by integrating flat plate bifacial PV cells [1–4]. Thus, the hybrid CPV/PV technology maximizes power output per unit area [5]. In fact, the hybrid bifacial CPV/PV module presented here reaches a world record efficiency of 34.2% [6, 7] for conversion of the reference AM1.5g spectrum at standard test conditions (STC). This hybrid module technology, named EyeCon, was developed by Fraunhofer ISE and employs III-V wafer-bonded (//) four-junction (4J) CPV cells (GaInP/GaAs//GaInAsP/GaInAs) [8, 9] to convert the direct light which is concentrated 321x by a silicone-on-glass Fresnel lens plate, as shown in **Figure 1a**. Underneath the CPV cells, the module integrates bifacial crystalline silicon (c-Si) solar cells that absorb the transmitted diffuse irradiance, direct light spilled by the lens and the albedo radiation that arrives at the rear side. Furthermore, CPV and flat plate PV cells remain electrically isolated by a dielectric thermal adhesive that enables heat dissipation through the c-Si PV cell [10]. In consequence, radial temperature profiles develop around the CPV cells as qualitatively shown in **Figure 1b**. Moreover, the 4J EyeCon module is a dual-axis tracked 4-terminal device where the CPV and flat plate PV circuits are independent. Further details about the module architecture and outdoor characterization can be found in [7, 11].

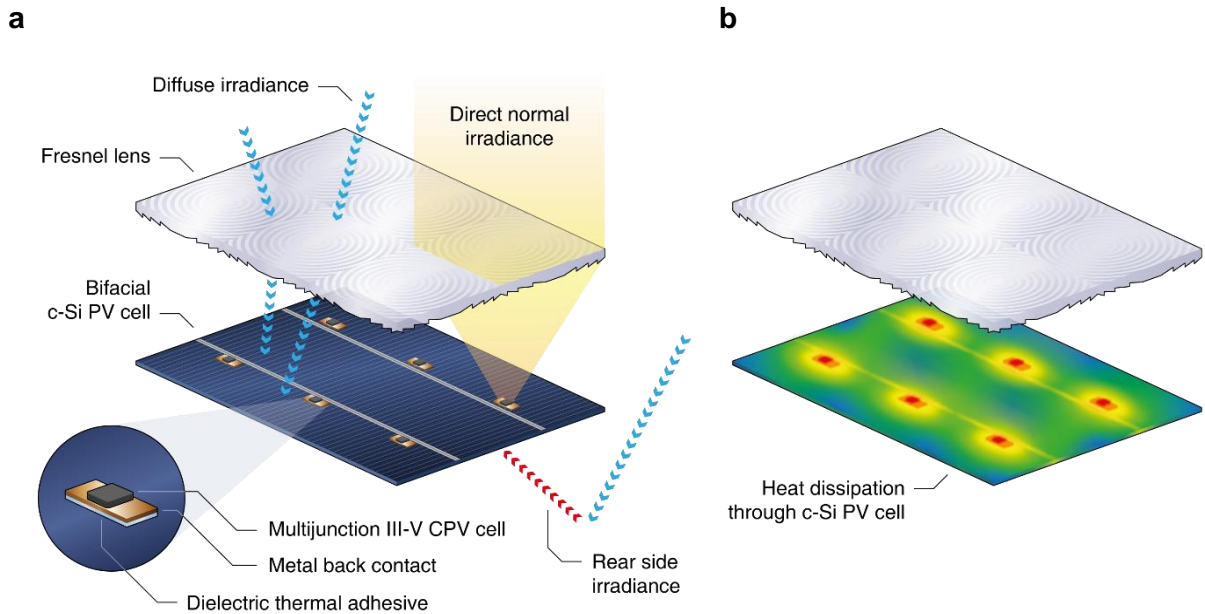

**Figure 1:** Diagram of the EyeCon, a bifacial hybrid CPV/PV module, explaining the architecture and functionalities of the device. Part **(a)** shows the concentration of direct sunlight onto the 4J III-V CPV cells by a Fresnel lens and the absorption of front and rear diffuse irradiance by the flat plate c-Si PV cell. Part **(b)** shows the temperature distribution on the c-Si PV cell as it effectively distributes the concentrated heat flux radially away from the CPV cells due to the use of a dielectric thermal adhesive. The CPV cells are mounted on metal back contacts to facilitate electrical interconnection.

Based on the outdoor characterization of the 4J EyeCon module, in this work, we independently modeled the power output of the CPV and flat plate PV arrays using empirical models [12, 13]. The functionality and performance of the models are discussed in Section 2. Then, in Section 3, the energy yield of the hybrid CPV/PV module is presented for numerous locations around the world using data of a typical meteorological year (TMY) as input parameters, i.e. global horizontal irradiance (GHI), direct normal irradiance (DNI), diffuse horizontal irradiance (DHI), ambient temperature ($T_{amb}$), atmospheric pressure ($P_{atm}$), relative humidity (RH) and wind speed ($V_{wind}$). Besides TMY data, the spectral parameters $Z_{1-2}$ and $Z_{1-3}$ for the 4J CPV cells [14] and the spectral mismatch (SMM) factor for the flat plate solar cells [15] are used to assess their response under an arbitrary solar spectrum. The method presented in reference [16] is used to estimate these spectral parameters from DNI, global normal irradiance (GNI), $P_{atm}$, precipitable water (PW) and air mass (AM). Also, in Section 3, the energy yield results of standard 4J CPV and c-Si flat plate PV technologies are presented for comparison. Fixed-tilt and single-axis tracked monofacial and bifacial configurations were considered for the analysis of flat plate PV technology. Then, the energetic advantage of hybrid CPV/PV devices over standard CPV and flat plate PV technologies is discussed in Section 4 in a geographical context. Finally, in Section 5 a techno-economic analysis based on assumptions regarding the system cost of each PV technology and their respective worldwide energy yield is presented.

## 2. Power Output Modeling

Calculating the power output of hybrid CPV/PV technology requires accurate modeling of multi-junction CPV and flat plate PV electrical behavior under varying spectral and meteorological conditions. Hence, two empirical models validated in the literature [12, 13] were used to perform such power output calculation. For both models, regression coefficients for the CPV and flat plate PV arrays of the 4J EyeCon module were derived in this work. The optical, thermal and spectral effects considered by these models are discussed in the following subsections along with their performance.

### *2.1 Multi-junction CPV Technology*

The multi-junction CPV power output model developed by Peharz et al. in [12] was used in this work for its simplicity and because its accuracy has been demonstrated in more than one location, i.e. Freiburg, Germany and Sevilla, Spain. The method is based on superposition and it separately models the short-circuit current responsivity ($I_{SC}$/DNI), the fill factor (FF) and the open-circuit voltage ($V_{OC}$) as the multi-linear regressions given in **Equations 1** to **3**, before calculating the maximum power output ($P_{CPV}$) as the product $P_{CPV} = I_{SC}/DNI \cdot FF \cdot V_{OC} \cdot DNI$.

$$\frac{I_{SC}}{DNI} = i_0 + i_1 \left|Z_{1-2} - Z_{1-2\_match}\right| + i_2 \left|Z_{1-3} - Z_{1-3\_match}\right| + i_3 \, T_{amb} + i_4 \, T_{amb}^{\;2} \qquad (1)$$

$$FF = f_0 + f_1 \left|Z_{1-2} - Z_{1-2\_match}\right| + f_2 \sqrt{\left|Z_{1-2} - Z_{1-2\_match}\right|} + f_3 \, T_{amb} + f_4 \, DNI \qquad (2)$$

$$V_{OC} = v_0 + v_1 \, T_{amb} + v_2 \, DNI + v_3 \, \ln(DNI) \qquad (3)$$

The inputs of the original model [12] consist of DNI, $T_{amb}$ and the spectral parameter $Z_{1-2}$, however, in this work we also included $Z_{1-3}$ in **Equation 1** to increase its robustness when using it in extremely dry or humid regions where light absorption by PW may introduce significant deviations. As it is explained in the appendix in subsection 7.1, using spectral parameters Z [17, 18] is a simple, yet effective way to assess the current mismatch of a 4J CPV cell when the solar spectrum shape varies due to the influence of aerosols scattering and water absorption. In this sense, $Z_{1-2}$ quantifies the impact of aerosols scattering on the relative photo-generation between sub-cells that absorb photons below 900 nm, whereas $Z_{1-3}$ quantifies the effect of scattering and water absorption between sub-cells that absorb photons below 600 nm and above 900 nm. Although the bottom two junctions of the 4J CPV cell absorb light beyond 900 nm, it is not necessary to introduce a third spectral parameter to differentiate them. This is because they only limit the current output under extremely humid conditions which are rarely encountered. Furthermore, $Z_{1-2\_match}$ and $Z_{1-3\_match}$ are the values where the optimum current-matching between the sub-cells occurs, whereas $Z_{min}$ corresponds to the smallest $Z_{1-2}$ value calculated from outdoor measurements. These values are given in **Table 1** for the CPV array of a 4J EyeCon and of a 4J FLATCON® [19] module, along with the regression coefficients $i_i$, $f_i$ and $v_i$ which weight the impact of each term.

In addition, we split the usage of **Equation 1** in two ranges ($Z_{1-2} \leq Z_{min}$ and $Z_{1-2} > Z_{min}$) because the measured $I_{SC}$/DNI of the 4J solar cells deviates from the modeled one for values below $Z_{min}$, as it is explained in the appendix in subsection 7.2. Similarly, we split the usage of **Equation 2** in two ranges. For $Z_{1-2} > Z_{1-2\_match}$, we use the coefficients derived from the measured FF beyond the current-matching condition, i.e. when one of the infrared absorbing sub-cells limits the current output. For $Z_{1-2} \leq Z_{1-2\_match}$, we use the coefficients derived from the measured FF when the GaInP sub-cell is limiting. However, given the fact that below $Z_{min}$ the GaInP sub-cell continues to limit, the FF is fixed at 90% for $-1 \leq Z_{1-2} \leq Z_{min}$ because its trend at $Z_{1-2} = Z_{min}$ already suggests that the FF flattens, as shown in **Figure 2b**. In the following, we demonstrate and explain the application of the model to the CPV array of the 4J EyeCon module.

It is well known that monolithic multi-junction solar cells are strongly sensitive to spectral variation despite the linear response of their sub-cells to irradiance. The reason is that the current output is limited by the junction with the lowest photo-generation, given that each sub-cell absorbs a distinct part of the solar spectrum. Moreover, the limiting sub-cell can change as the solar spectrum varies.

As shown in **Figure 2a**, the $I_{SC}$/DNI ratio of the 4J CPV array reaches a maximum when the sub-cells are current-matched. In this case that occurs when the spectral parameters $Z_{1-2}$ = -0.027 and $Z_{1-3}$ = -0.017. It is important to note that in a first approximation $I_{SC}$/DNI linearly increases with $Z_{1-2}$ and $Z_{1-3}$ until current matching occurs. This is because the limiting junction (GaInP) which absorbs light in the visible range gains current as the solar spectrum shifts towards shorter wavelengths. However, $I_{SC}$/DNI decreases linearly beyond the current-matched condition when one of the infrared absorbing sub-cells begins to limit [20].

**Table 1:** Multi-linear regression coefficients of the CPV power output model of a 4J hybrid CPV/PV EyeCon ($\eta_{STC}$ = 34.2%) and of a conventional 4J CPV FLATCON® ($\eta_{CSTC}$ = 36.7%) module with their respective normalized root mean square errors (NRMSE).

| Coefficient | Unit | 4J CPV array (hybrid) | | 4J CPV array (conventional) | |
|---|---|---|---|---|---|
| $Z_{1\text{-}2\_match}$ | abs. | -0.027 | | -0.030 | |
| $Z_{1\text{-}3\_match}$ | abs. | -0.017 | | -0.041 | |
| $Z_{min}$ | abs. | -0.23 | | -0.28 | |
| | | $Z_{1\text{-}2} \leq Z_{min}$ | $Z_{1\text{-}2} > Z_{min}$ | $Z_{1\text{-}2} \leq Z_{min}$ | $Z_{1\text{-}2} > Z_{min}$ |
| $i_0$ | mA/(W/m$^2$) | 0.904 | 0.876 | 0.617 | 0.694 |
| $i_1$ | mA/(W/m$^2$) | -0.034 | -0.295 | -0.026 | -0.615 |
| $i_2$ | mA/(W/m$^2$) | -0.894 | -0.836 | -0.623 | -0.257 |
| $i_3$ | μA/(K·W/m$^2$) | 9.01 | 9.01 | 1.797 | 1.797 |
| $i_4$ | μA/(K$^2$·W/m$^2$) | -0.186 | -0.186 | -0.018 | -0.018 |
| **NRMSE** | % | **5.4** | **1.5** | **5.6** | **1.2** |
| | | $Z_{1\text{-}2} \leq Z_{1\text{-}2\_match}$ | $Z_{1\text{-}2} > Z_{1\text{-}2\_match}$ | $Z_{1\text{-}2} \leq Z_{1\text{-}2\_match}$ | $Z_{1\text{-}2} > Z_{1\text{-}2\_match}$ |
| $f_0$ | % | 71.77 | 82.47 | 81.65 | 83.52 |
| $f_1$ | % | 16.44 | 24.36 | -22.20 | 51.52 |
| $f_2$ | % | 23.93 | -1.702 | 24.03 | -3.452 |
| $f_3$ | %/K | -0.008 | -0.079 | -0.017 | 0.042 |
| $f_4$ | %/(W/m$^2$) | 0.0104 | 2.37e-4 | -0.001 | -0.003 |
| **NRMSE** | % | **1.3** | **0.4** | **0.6** | **0.5** |
| $v_0$ | V | 48.309 | | 36.715 | |
| $v_1$ | mV/K | -88.2 | | -120 | |
| $v_2$ | mV/(W/m$^2$) | -3.22 | | -5.13 | |
| $v_3$ | V | 0.356 | | 3.037 | |
| **NRMSE** | % | **0.5** | | **0.7** | |

**Figure 2b** shows the opposite effect for the FF that exhibits a minimum under current-matched conditions and increases asymmetrically away from them as expected from previous observations [20]. Scattering in **Figure 2b** originates from FF dependencies on cell and lens temperature as well as on DNI intensity which causes series resistance effects.

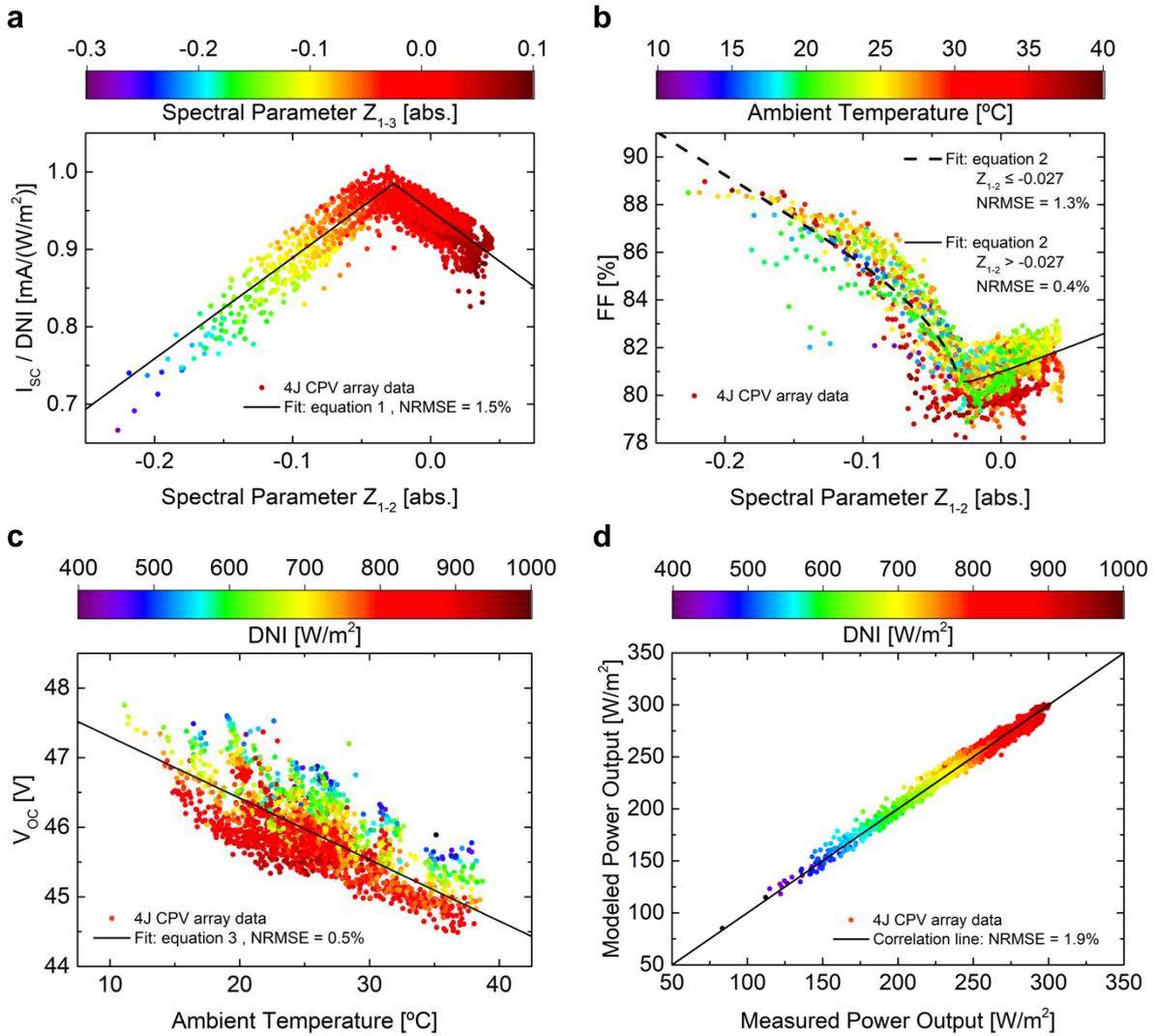

**Figure 2:** Measured **(a)** $I_{SC}$/DNI versus spectral parameter $Z_{1-2}$ with $Z_{1-3}$ color coding, **(b)** FF versus spectral parameter $Z_{1-2}$ with $T_{amb}$ color coding, **(c)** $V_{OC}$ versus $T_{amb}$ with DNI color coding and **(d)** correlation plot of modeled versus measured power output per unit area of the 4J CPV array of the EyeCon module with DNI color coding. The legends display references to the equations and normalized root mean square errors of the power output model [12].

In addition, concentration logarithmically boosts $V_{OC}$ according to DNI, but it also emphasizes a reduction in voltage when the cell temperature increases with $T_{amb}$ and DNI as shown in **Figure 2c**. Also, temperature leads to lens deformation and refractive index variation that influence the current generated by the concentrator cells [21, 22]. Here it is important to mention that all the temperature effects are considered without explicitly calculating the CPV cell temperature, but rather through $T_{amb}$ and DNI which implicitly determine it. For reference, the mean cell temperature of the 4J CPV array during the outdoor measurements was (69 ± 9) ºC.

**Figure 2d** shows a correlation between the modeled power output of the CPV array of the 4J EyeCon module and the 2117 measurements performed in Freiburg, Germany during the summer of 2019. The measurements were performed over 43 different days and cover a wide range of meteorological and spectral conditions, as portrayed in **Figures 2a** to **2c**. As denoted by a low normalized root mean square error (NRMSE) of 1.9%, the empirical model [12] shows a similar level of accuracy as the software PVsyst [23], that uses the one-diode model, for which a NRMSE of 3.7% was reported for the Concentrix CPV technology [24, 25]. In this work, NRMSE were normalized with the mean value of the data in all cases. Here it is important to mention that the accuracy of our CPV power output model was tested with instantaneously measured DNI, whereas the Concentrix CPV model was validated using hourly averaged DNI. Hence, we expect the effect of using hourly averaged DNI from TMY data to have a low impact in the accuracy of our CPV power output model.

## *2.2 Single-junction Flat Plate PV Technology*

Relative to multi-junction devices, single p-n junction solar cells are simpler to model due to their less complex architecture. Particularly, c-Si PV cells have been extensively studied and thus, their outdoor performance under varying conditions is well understood [26]. Moreover, they typically respond linearly to irradiance and temperature changes which are the dominant effects that determine their power output [27]. In this work, we used the empirical model defined in [13] because it accurately considers these effects and it is less computationally intensive as the widely used one-diode model [23]. Furthermore, it was demonstrated that it yields a low NRMSE of 1.2% for 18 different mono- and multicrystalline Si PV modules under one sun illumination. Thus, it has been used in several PV energy yield studies over large geographical areas [28–30]. Its mathematical expression is given in **Equation 4**,

$$P_{PV} = G\,(p_0 + p_1\,ln(G) + p_2\,ln(G)^2 + p_3\,T + p_4\,T\,ln(G) + p_5\,T\,ln(G)^2 + p_6\,T^2) \qquad (4)$$

where the power output at the maximum power point is multi-linearly regressed as a function of the normalized effective irradiance, $G$ = SMM · ($G_{front}$ + φ · BTI) / $G_{ref}$, and of the cell temperature gradient, $T = T_{cell} - 25°C$, according to the $p_i$ weighting coefficients. Therefore, the model considers the effect of bifaciality by scaling the backplane tilted irradiance (BTI) with the bifaciality factor (φ) of the module. The spectral variation is considered by multiplying the front irradiance ($G_{front}$) plus the scaled BTI with the spectral mismatch factor (SMM) defined in the standard for PV devices (IEC 60904-7 [15]), as demonstrated by the author of the power output model [13] in [28]. The explanation on how to calculate the spectral mismatch factor following the approach in [16] can be found in the appendix in subsection 7.1. For a bifacial one sun PV module G = SMM · (GTI + φ·BTI) / 1000 W/m², whereas for the bifacial flat plate PV array inside the hybrid EyeCon module G = SMM · (DTI + φ · BTI) / 100 W/m² because it only receives diffuse irradiance. Here GTI and DTI stand for global and diffuse tilted irradiance, whereas $G_{ref}$ = 1000 and 100 W/m² for a one sun and a for a hybrid module, respectively. In the case of a conventional PV module, we used a bifaciality factor of 0.9 because this is a typical value for state-of-the-art devices [31]. For the Si PV array of the EyeCon module we used its measured value from [7], i.e. φ = 0.64. As shown in **Figure 3a**, the power output model [13] also performs well (NRMSE = 3.3%) under partial illumination, since the bifacial flat plate PV cells in the EyeCon module only receive diffuse light.

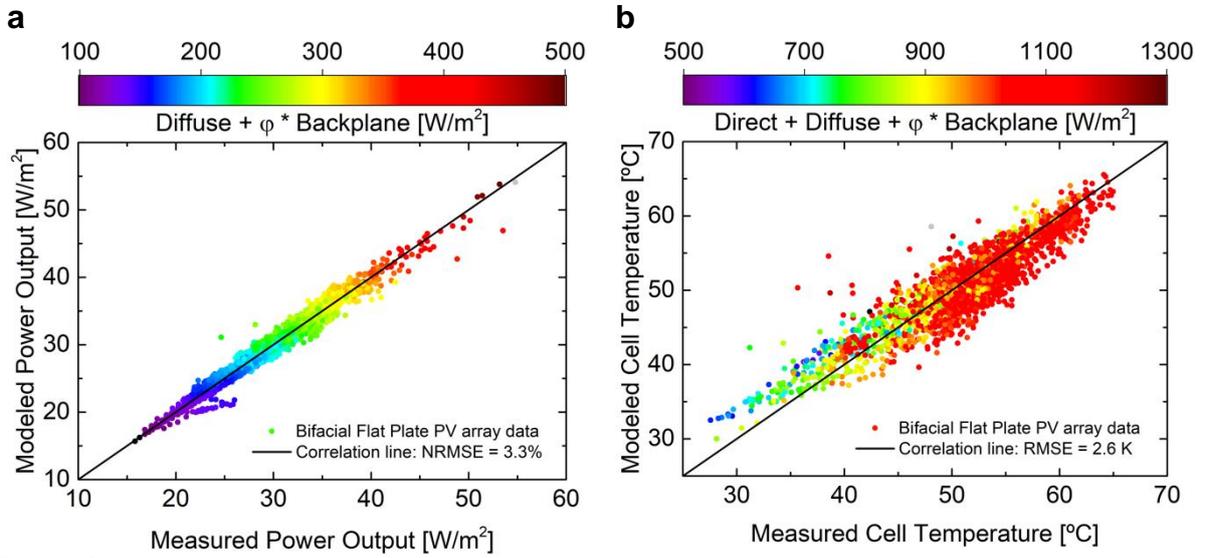

**Figure 3:** Correlation plots of modeled versus measured **(a)** power output per unit area color coded according to the optically effective irradiance and **(b)** cell temperature of the flat plate c-Si PV cells inside the EyeCon module color coded according to the thermally effective irradiance. The legends display normalized and absolute root mean square errors, respectively.

Moreover, the temperature of the flat plate c-Si PV cells ($T_{PV}$) is calculated using the model in the PV energy rating standard (IEC 61853-2 [32]) which is described in [33]. However, the measured temperature of the Si cell array inside the EyeCon module was calculated with its $V_{OC}$ and temperature coefficient using the method described in the IEC 60904-5 standard [34]. As we previously showed in [10], this method works well since it yields the effective cell temperature based on electrical characteristics despite the inhomogeneous temperature profile on the Si cells. As given in **Equation 5**, the inputs of the cell temperature model [33] are $T_{amb}$, $V_{wind}$ and the absorbed irradiance ($G_{abs}$),

$$T_{PV} = T_{amb} + \frac{G_{abs}}{u_0 + u_1 V_{wind}} \tag{5}$$

where $u_0$ and $u_1$ are the regression coefficients that represent the steady-state and the wind dependent heat exchange coefficients, respectively. For a conventional flat plate PV module, $G_{abs}$ = GTI + φ · BTI, whereas for the bifacial PV array of the hybrid EyeCon module $G_{abs}$ = DNI + DTI + φ · BTI, because the Si cells also dissipate the DNI absorbed by the CPV cells. **Figure 3b** shows how the cell temperature model [33] estimates the temperature of the c-Si PV cells with a low RMSE of 2.6 K.

Therefore, we chose to use both of these models [13, 33] in our worldwide calculation because they reproduce our measurements with sufficient accuracy. In addition, they address the main irradiance, temperature and spectral effects that affect the performance of single-junction flat plate PV technology. Their regression coefficients are given in **Table 2** with their respective NRMSE and RMSE.

**Table 2:** Multi-linear regression coefficients of the flat plate PV power output and cell temperature models of a 4J hybrid CPV/PV ($\eta_{STC}$ = 34.2%) and of a flat plate c-Si PV ($\eta_{STC}$ = 20%) module with their respective NRMSE and RMSE.

| Coefficient | Unit | c-Si PV array (hybrid) | c-Si PV array (conventional) |
|---|---|---|---|
| φ | abs. | 0.64 | 0.90 |
| $p_0$ | W/m² | 16.9 | 200 |
| $p_1$ | W/m² | 0.007 | -3.448 [13] |
| $p_2$ | W/m² | -1.006 | -8.094 [13] |
| $p_3$ | W/(m²·K) | -0.093 | -0.940 [13] |
| $p_4$ | W/(m²·K) | -0.174 | 0.030 [13] |
| $p_5$ | W/(m²·K) | 0.070 | 0.029 [13] |
| $p_6$ | W/(m²·K²) | 0.002 | 0.001 [13] |
| **NRMSE** | **%** | **3.3** | **1.2** [13] |
| $u_0$ | W/(m²·K) | 37.8 | 25.0 [33] |
| $u_1$ | W·s/(m³·K) | 0.95 | 6.84 [33] |
| **RMSE** | **K** | **2.6** | **1.9** [33] |

The $p_i$ coefficients of the conventional c-Si PV array correspond to those given in [13] multiplied by $P_{STC}$, i.e. 200 W/m².

## 3. Worldwide Annual Energy Yield Estimation

Estimating the worldwide energy yield of hybrid CPV/PV technology requires reliable spectral and meteorological data as input for the power output models described in the previous section. For our analysis, we used hourly TMY data with a spatial resolution of 0.25° × 0.25° (ca. 28 km × 28 km). TMY files contain the most representative GHI, DHI, DNI, $T_{amb}$, RH, $P_{atm}$ and $V_{wind}$ values of every hour of the year for each location [35]. The data is derived from observations of geostationary meteorological satellites and is freely available online from PVGIS [36–38]. As displayed in **Figure 4**, almost all populated areas of the world are covered except for parts of North and South America, north and eastern Russia, eastern China and most of Australia. In total 8760 power output calculations were performed for each of the 155 687 locations available from PVGIS. Additionally, twelve monthly-average ground albedo maps from 2015 were used to create one yearly-average map to calculate the irradiance reflected off the ground. The albedo maps are freely available online from NASA Earth Observations for 2015 and other years [39]. For comparison of the hybrid CPV/PV potential against its independent counterparts, we also estimated the energy yield of conventional dual-axis tracked 4J CPV and monofacial or bifacial flat plate PV modules in a fixed-tilt orientation at the optimum tilt angle or with single-axis tracking. Given that single- and dual-axis tracked systems are rarely mounted on rooftops, we assumed ground level mounting for all technologies. For all calculations, a single 1 m × 2 m module in portrait orientation was considered to keep the analysis independent of system size and field configuration. This simplification entails that row-to-row shading losses are not considered in our calculations. According to [40], these are in the range of 10% for single- and dual-axis tracked systems for ground coverage ratios (GCR) of 30 and 20%, whereas for fixed-tilt installations this occurs well-beyond a GCR of 50%. Therefore, the larger land requirement of tracked PV systems should be considered when using the following energy yield results to analyze a particular case. Furthermore, the potential current mismatch between bifacial Si cells within a flat-plate PV module due to shading by the mounting structure is not considered in our calculations. From these perspectives, our results illustrate an upper boundary scenario for the performance of every investigated technology. Moreover, estimating the effective irradiance on the front and rear sides of the fixed-tilt, single- or dual-axis tracked plane is a complex endeavor that requires several optical considerations and geometrical assumptions [41]. A detailed description of the module orientations, the tracking specifications and the algorithms used for the calculation of GTI, DTI and BTI is given in the appendix in subsection 7.3.

### *3.1 Dual Axis-Tracked Hybrid III-V 4J CPV/PV c-Si Bifacial Technology*

The energy yield calculation of hybrid CPV/PV technology uses the power output models of multi-junction CPV [12] and single-junction flat plate PV [13] modules fitted to the measured data of the 4J EyeCon module as described in detail in subsections 2.1 and 2.2. The regression coefficients are given in **Tables 1** and **2**, whereas the worldwide energy yield map of hybrid CPV/PV technology is shown in **Figure 4a**. Given the fact that CPV is the main efficiency driver of the hybrid module, its energy yield is highest in locations with abundant DNI. These locations, where the mean annual yield is (982 ± 41) kWh/m$^2$ and up to 1150 kWh/m$^2$, are shown in red and dark red. Moreover, they are typically characterized by yearly DNI above 2500 kWh/m$^2$, low AOD and PW, high average $T_{amb}$ and arid climate. Nevertheless, the bifacial flat plate PV array generates in average 15.6% of the total energy of the hybrid module in these places. There, the missing energy yield calculations due to the lack of TMY data for countries like Chile and Australia are expected to be in the same range as for South Africa because they share similar latitude, spectral and meteorological conditions.

Considering the whole world, the 4J CPV cells convert in average (30.2 ± 1.3) % of the yearly DNI, whereas the bifacial Si cells harvest (14.1 ± 1.2) % and (7.1 ± 1.1) % of the annual diffuse and backplane tilted irradiance, respectively. This corresponds to an average conversion of the total front and rear solar resource on a dual-axis tracked plane of (22.6 ± 1.4) %. Additionally, to quantify the impact of neglecting spectral variation, we calculated the energy yield in the same manner but with constant $Z_{1-2} = Z_{1-3} = 0$ and SMM = 1 throughout the year for all locations. The result was an average overestimation of the CPV and flat plate PV yields by (6.2 ± 4.1) %$_{rel}$ and (2.1 ± 1.5) %$_{rel}$, which equates to a combined hybrid yield overestimation of (5.1 ± 3.1) %$_{rel}$ when the variation of the solar spectrum is ignored. As expected, the largest impact is on multi-junction solar cells since they are more sensitive to spectral variation due to the current-mismatch situations limiting the flow of current through the device. Moreover, the magnitude of the effect is between the values reported in [42] for Europe, Africa and the Middle East, i.e. 2 - 4%$_{rel}$, and in [43] for 47 different locations along the American continent, i.e. 12.4%$_{rel}$.

## 3.2 Dual-Axis Tracked III-V 4J CPV Technology

The energy yield calculation of standard CPV technology uses the same power output model as the one used for the CPV part in subsection 2.1. However, the fit parameters are different as they have been derived from measured data of a 4J FLATCON® CPV module with a 36.7% [19] efficiency at concentrator standard test conditions (CSTC). The regression coefficients are given in **Table 1**. The module is equipped with the same type of 4J CPV solar cells as the EyeCon module, but these are mounted on a metal heat spreader made of Cu. Additionally, it uses smaller Fresnel lenses with 16 cm$^2$ and achieves a lower geometrical concentration of 226x. Therefore, the 4J CPV cells of the FLATCON® module are expected to operate 20 K lower as in the EyeCon module [10, 11]. The map of the worldwide energy yield for CPV technology is shown in **Figure 4b**. In ideal CPV locations where yearly DNI exceeds 2500 kWh/m$^2$, the technology delivers in average (894 ± 40) kWh/m$^2$ annually, which is significantly lower than the yield of the total hybrid system. Comparing the pure CPV performance, this value is 8% higher than the energy generated by the 4J CPV cells alone in the hybrid module, which is attributed to the higher FLATCON® CSTC efficiency and its lower operating temperature. Moving away from high DNI regions, the energy yield of CPV rapidly falls into a similar range as that of flat plate c-Si PV technology. Nevertheless, considering the whole world, the conventional 4J FLATCON® module is able to convert in average (32.6 ± 1.1) % of the yearly DNI into electricity.

## 3.3 Fixed-Tilt or Single-Axis tracked Monofacial or Bifacial c-Si PV Technology

The energy yield calculation of conventional c-Si PV technology in this work uses the same power output model as the one used for the flat plate PV part of the hybrid CPV/PV module in subsection 2.2. However, we used the regression coefficients for a generic c-Si module derived by Huld et al, in [13]. In addition, we employed representative efficiency and bifaciality values of state-of-the-art c-Si PV cells, i.e. $\eta_{STC}$ = 20% and $\varphi$ = 0 (for monofacial) or $\varphi$ = 0.9 (for bifacial), for our calculations. The regression coefficients are given in **Table 2** and the worldwide energy yield maps for flat plate c-Si PV technology are shown in **Figures (4c)** for bifacial single-axis tracked, **(4d)** for monofacial single-axis tracked, **(4e)** for bifacial fixed-tilt and **(4f)** for monofacial fixed-tilt. First, flat plate PV also delivers its highest annual yields in the highest insolated regions of the world as CPV does. As a baseline, monofacial fixed-tilt PV generates (454 ± 15) kWh/m$^2$ in these high GHI locations, shown in green in **Figure 4f**, where the irradiation is above 1900 kWh/m$^2$. In mid GHI regions typically located at latitudes beyond 45º where the irradiation is around (1110 ± 201) kWh/m$^2$, the yield drops to (251 ± 44) kWh/m$^2$.

Exchanging monofacial for bifacial PV modules in high GHI locations represents a yield increase calculated in this work of (5.3 ± 1.4) %$_{rel}$, whereas in more northern latitudes a larger gain of (8.5 ± 2.2) %$_{rel}$ is calculated. The calculated bifacial gain is lower in high GHI locations because the clearer atmosphere and its lower yearly AM favor front side power generation [44]. If instead, monofacial PV output is enhanced with single-axis tracking, the larger solar resource collected on the front plane of the array boosts the energy yield by (33.4 ± 1.6) %$_{rel}$ and (29.7 ± 4.0) %$_{rel}$ in high and mid GHI locations. These results are in accordance with the findings in [29]. Combining bifaciality with single-axis tracking improves flat plate PV performance by (46.9 ± 3.9) %$_{rel}$ and (44.5 ± 5.5) %$_{rel}$ in high and mid GHI locations. This is significantly higher than the superposition of both independent gains. The reason is that lower ground self-shading improves rear light collection when the module is elevated 1 m above the ground as we assumed in the single-axis tracked calculation [44, 45]. Considering the whole world, each flat plate PV configuration converts in average (18.3 ± 0.7) % of the available annual irradiation, i.e. GTI for monofacial and GTI + BTI for bifacial modules. Furthermore, we investigated the influence of neglecting spectral variation and found that an average yield overestimation of (1.2 ± 1.1) ‰$_{rel}$ occurs for every flat plate PV configuration. The magnitude of this effect is slightly in the opposite direction of values reported in several other studies [28, 46–48] that found annual yield underestimations between 0 and 2%$_{rel}$ over Africa, Europe and Asia.

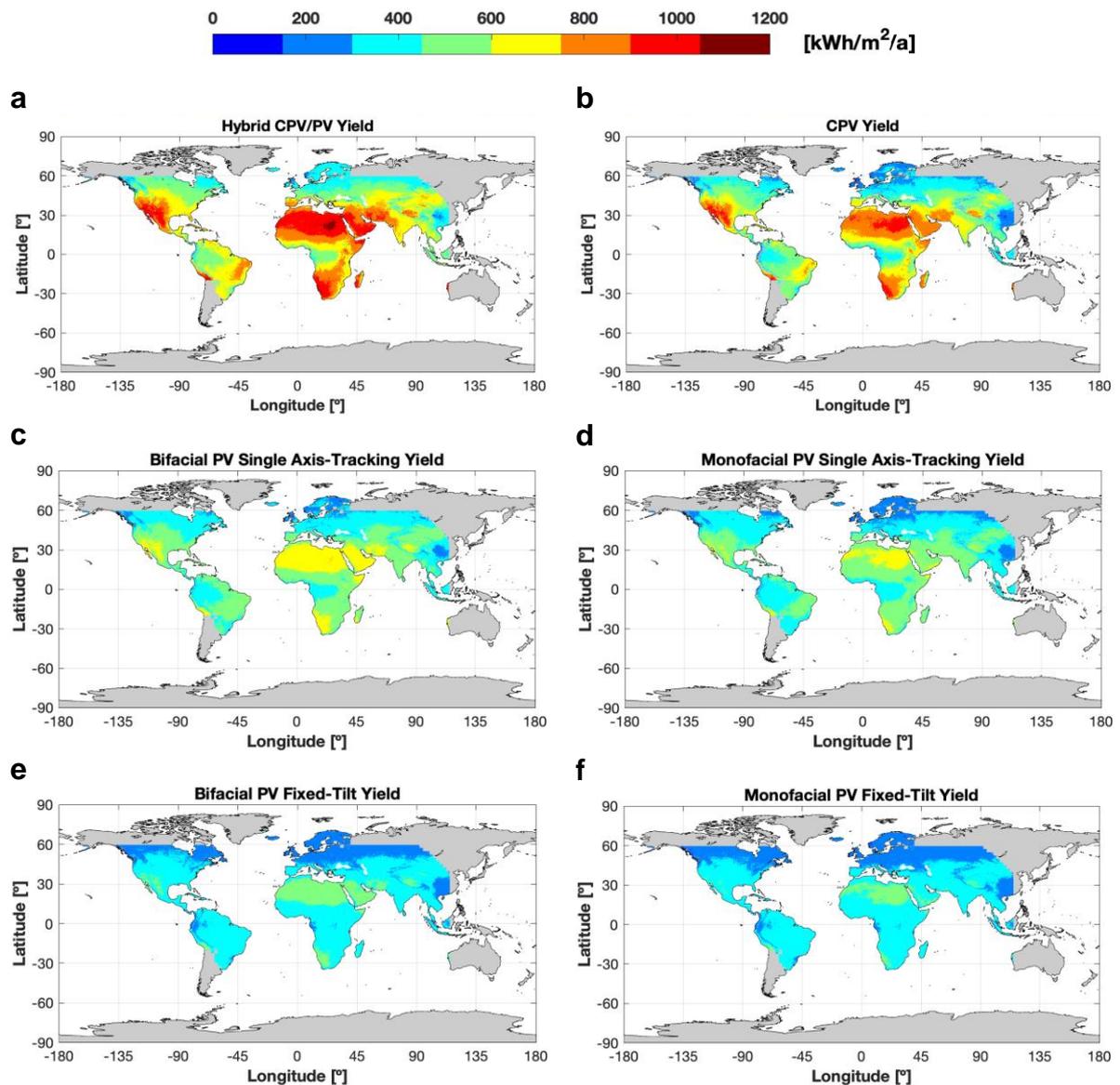

**Figure 4:** Calculated annual energy yield maps of **(a)** hybrid CPV/PV **(b)** CPV, **(c)** bifacial PV with single-axis tracking **(d)** monofacial PV with single-axis tracking, **(e)** bifacial PV fixed-tilt and **(f)** monofacial PV fixed-tilt technologies.

As a summary, **Table 3** compiles the mean annual energy yields and energy harvesting efficiencies calculated for each PV technology investigated plus the monofacial version of hybrid CPV/PV technology which is not shown in **Figure 4**. Comparing the mean energy yields with each other gives a general idea of the incremental gains that can be expected from increasing the technological complexity. Moreover, the mean energy harvesting efficiencies calculated for flat plate PV and CPV modules deviate from their nominal values by around 10%, whereas the discrepancy increases to 38% for hybrid CPV/PV technology. The reason is the strong dependence of its performance on the diffuse irradiance fraction. Therefore, a simple but naive calculation assuming only the efficiency at STC would yield disproportionately higher results in favor of the hybrid approach. Hence, the extensive modeling developed in this work is certainly valuable to establish a more accurate analysis.

**Table 3:** Summary of the annual energy yields (mean ± std. dev.) and comparison of the energy harvesting efficiencies of every investigated PV technology with respect to their efficiency at STC. In addition, the results for the monofacial version of hybrid CPV/PV technology are given.

| Technology | Annual Energy Yield [kWh/m$^2$] | Energy Harvesting Efficiency [%] | STC Efficiency Front / Rear [%] | Relative Efficiency Deviation [%] |
|---|---|---|---|---|
| Monofacial Fixed-Tilt PV | 340 ± 77 | 18.4 ± 0.7 | 20 / 0 | 8.7 |
| Bifacial Fixed-Tilt PV | 357 ± 78 | 18.3 ± 0.6 | 20 / 18 | 8.8 |
| Monofacial Single-Axis Tracked PV | 437 ± 105 | 18.4 ± 0.7 | 20 / 0 | 8.7 |
| Bifacial Single-Axis Tracked PV | 478 ± 115 | 18.1 ± 0.7 | 20 / 18 | 9.5 |
| Conventional CPV | 556 ± 195 | 32.6 ± 1.1 | 36.7 / 0 | 12.6 |
| Hybrid CPV/PV (monofacial) | 631 ± 186 | 24.8 ± 1.5 | 34.2 / 0 | 37.9 |
| Hybrid CPV/PV (bifacial) | 655 ± 191 | 22.6 ± 1.4 | 34.2 / 11 | 38.8 |

Energy harvesting efficiency = annual energy yield / accessible solar resource for each technology.

## 4. Hybrid CPV/PV Energy Yield Advantage

**Figure 5** geographically shows the advantage in electrical energy production of using hybrid CPV/PV technology with respect to its individual counterparts. Most importantly, hybrid CPV/PV generates more energy in over 99.9% of the locations investigated. As shown in **Figures 5a** to **5d**, hybrid CPV/PV delivers in average 116, 105, 62 and 47% more energy per m$^2$ of module area relative to monofacial fixed-tilt, bifacial fixed-tilt, monofacial single-axis tracked and bifacial single-axis tracked PV technologies in locations with yearly DNI above 2500 kWh/m$^2$ (i.e. the locations shown in red and dark red in **Figure 4a**). In comparison with CPV, **Figure 5e** shows that the annual energy yield of hybrid CPV/PV technology is in average 20.8% higher. However, as opposed to the comparison with flat plate PV, hybrid CPV/PV generates the least energy surplus relative to CPV in high DNI regions, i.e. 9.8% in average.

These opposite trends suggest that there should be a set of optimum locations with particular meteorological conditions where hybrid CPV/PV technology presents its largest advantage over conventional CPV and flat plate PV technologies.

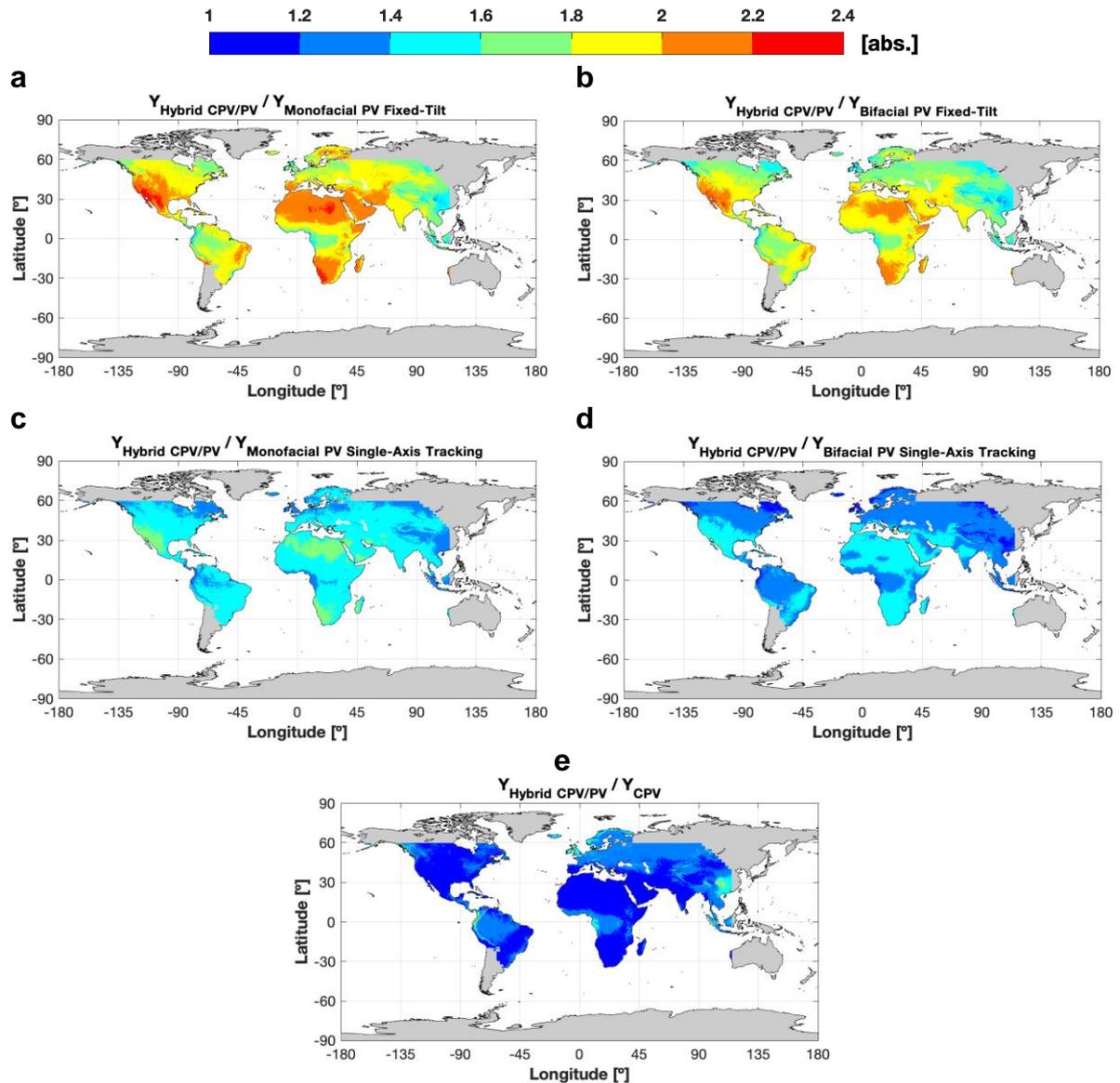

**Figure 5:** Energy yield ratio maps of hybrid CPV/PV technology over **(a)** monofacial PV fixed-tilt, **(b)** bifacial PV fixed-tilt, **(c)** monofacial PV with single-axis tracking, **(d)** bifacial PV with single-axis tracking and **(e)** CPV.

Therefore, in **Figure 6a**, we plot the energy yield ratio between hybrid CPV/PV technology ($Y_{Hybrid\ CPV/PV}$) and its closest contender, i.e. $Y_{closest} = \max(Y_{CPV}, Y_{Bifacial\ PV\ Single-Axis\ Tracking}, Y_{Monofacial\ PV\ Single-Axis\ Tracking}, Y_{Bifacial\ PV\ Fixed-Tilt}, Y_{Monofacial\ PV\ Fixed\ Tilt})$, for which $Y_{closest} = Y_{CPV}$ in 80% of the locations and $Y_{closest} = Y_{Bifacial\ PV\ Single-Axis\ Tracking}$ in the remaining 20%. Displayed in red, hybrid CPV/PV technology offers an advantage in electrical energy production beyond 30% over the runner-up PV technology in areas where the average yearly DNI and GHI are in the range of (1152 ± 158) kWh/m$^2$ and (1494 ± 365) kWh/m$^2$, respectively.

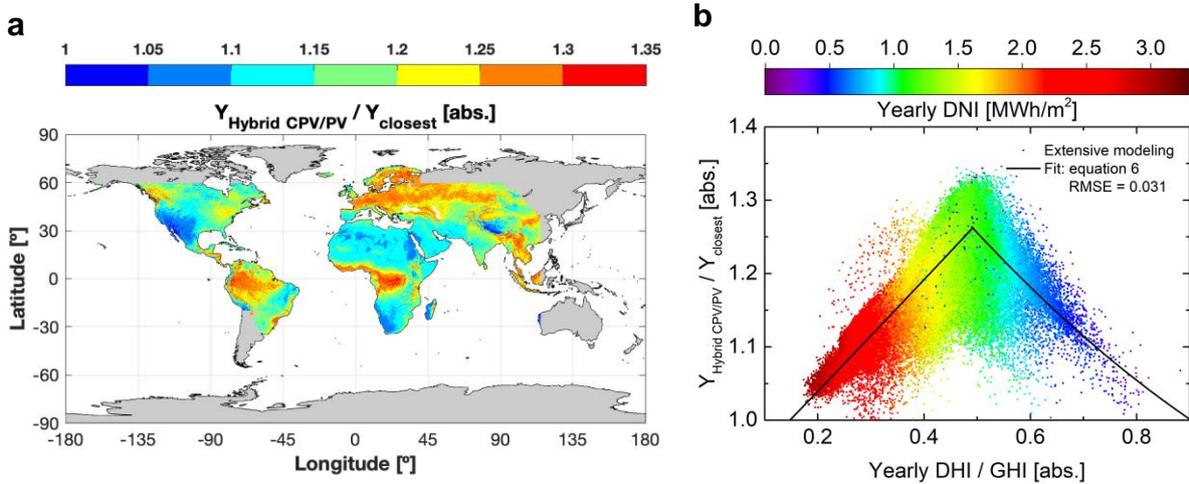

**Figure 6:** Energy yield ratio map **(a)** of hybrid CPV/PV technology over its closest contender, i.e. CPV and bifacial PV single-axis tracked modules: $Y_{closest} = \max(Y_{CPV}, Y_{Bifacial\ PV\ Single-Axis\ Tracking})$. Scatter plot **(b)** of the yield ratio shown in (a) as a function of annual DHI/GHI ratio and color coded according to yearly DNI in MWh/m$^2$. Each point in the scatter plot (b) corresponds to a distinct location in map (a).

Furthermore, **Figure 6b** shows the same yield ratio plotted against yearly DHI/GHI ratio. The result is a roof-top shaped function with a maximum value at DHI/GHI = 0.49. The left slope corresponds to $Y_{Hybrid\ CPV/PV} / Y_{CPV}$ and shows that as the diffuse irradiance fraction increases, the yield of the hybrid module relative to CPV increases up to 26%. This is because the bifacial flat plate PV cells integrated in the hybrid module additionally convert the diffuse light into electricity. Beyond that point, the right slope, i.e. $Y_{Hybrid\ CPV/PV} / Y_{Bifacial\ PV\ Single-Axis\ Tracking}$, decreases as the yearly DHI/GHI ratio increases. The reasons are that the energy generated by the high efficiency CPV cells decreases while the bifacial flat plate PV cells inside the hybrid module cannot compensate this deficit due to their lower efficiency and the higher reflection losses of diffuse irradiance. Thus, hybrid CPV/PV technology outperforms in average conventional CPV modules by 10 %$_{rel}$ in high DNI locations (DHI/GHI < 0.3) and bifacial PV single axis-tracked modules by 16%$_{rel}$ in regions with extremely cloudy conditions (DHI/GHI > 0.6). In locations with average hazy conditions (0.45 < DHI/GHI < 0.55) hybrid CPV/PV technology finds its optimal performance and generates between 10 and 34% more energy than its counterparts with an average of (25 ± 4.3) %$_{rel}$.

In an attempt to calculate the energy yield advantage of hybrid CPV/PV technology in a simpler manner, we regressed $Y_{Hybrid\ CPV/PV} / Y_{closest}$ over yearly DHI/GHI ratio and yearly DNI. We chose these two quantities because they gave the lowest RMSE and the highest adjusted coefficient of determination (adj-R$^2$) out of all the TMY parameters. The resulting expression is given in **Equation 6**.

$$\frac{Y_{Hybrid\ CPV/PV}}{Y_{closest}} = \begin{cases} 0.865 + 0.798\ \frac{DHI}{GHI} - 0.005\ DNI & ; \quad \frac{DHI}{GHI} < 0.49 \\ 1.471 - 0.542\ \frac{DHI}{GHI} - 0.050\ DNI & ; \quad \frac{DHI}{GHI} \geq 0.49 \end{cases} \quad (6)$$

Despite a modest RMSE of 3.1%, the bias of **Equation 6** is large enough to significantly overestimate the low gains while underestimating the large ones. Nevertheless, it serves as a first glance approach for projecting the potential of hybrid CPV/PV technology in a worldwide context. For example, in a high DNI location, like León, Mexico, (yearly DNI of 2425 kWh/m$^2$) where hybrid CPV/PV would generate 10.9% more energy than CPV, **Equation 6** predicts 10%. On the other hand, it predicts a 22.8% higher yield relative to bifacial PV single-axis tracking in a mid GHI location, like Ingolstadt, Germany, (yearly GHI of 1096 kWh/m$^2$) where our extensive modeling estimated 26.2%.

## 5. Techno-Economic Analysis of Hybrid CPV/PV Technology

The most important metric regarding the economics of a PV technology is cost of electricity. To round up our analysis we discuss here economic advantages for the hybrid CPV/PV technology in different parts of the world. For this we calculated the relative electricity cost of the hybrid CPV/PV technology with respect to its closest competitor, i.e. a conventional CPV or a bifacial single-axis tracked PV system. To do so, we need to make two assumptions. The first assumption is that a bifacial PV single-axis tracked system is 14% more expensive than its monofacial fixed-tilt counterpart. According to [49], 14% is a mean value resulting from a worldwide cost assessment of different flat-plate c-Si PV technologies. Second, we assume that a hybrid CPV/PV system is more expensive than a conventional CPV system by 11% of the cost of a monofacial fixed-tilt PV installation. This assumption is based on the fact that the average cost of a c-Si PV cell ($0.104/W$_p$) [50] corresponds to 11% of the average cost of a utility scale (100 MW) fixed-tilt PV installation ($0.94/W$_p$) [51]. We do not account for extra costs associated with the interconnection and lamination of the c-Si PV cells inside the hybrid CPV/PV module as this is approximately compensated by a reduced Ag content on the Si cells (lower irradiance requires less metal) and the removal of the metal heatsink material from the conventional CPV module.

As explained in more detail in the appendix in subsection 7.4, combining these cost correlations with the energy yield maps in **Figures 4a**, **4b** and **4c** enables the calculation of an electricity cost ratio between the hybrid CPV/PV technology and its closest competitor as a function of the ratio between the cost of a conventional CPV system and a monofacial fixed-tilt PV system. As an example, in **Figure 7** we have applied this method to the case of a CPV system being in total 50% more expensive than a fixed-tilt PV system. The authors believe that this is a reasonable assumption which is achievable today for production volumes on the order of hundred MW/year. Nevertheless, the same method can also be used in the future assuming other cost ratios between CPV and conventional PV technology.

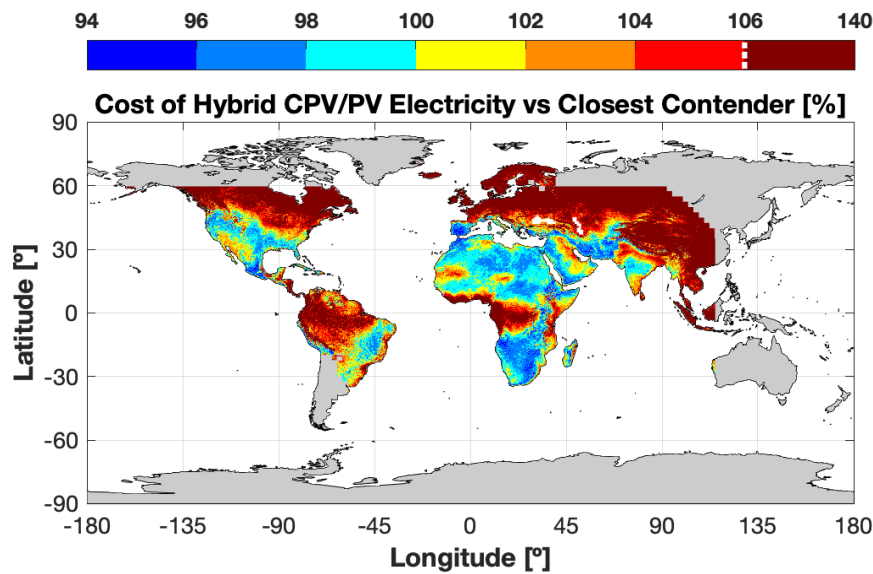

**Figure 7:** World map showing the regions (blue tones) where the hybrid CPV/PV technology is expected to have an overall cost advantage compared to its closest contender, i.e. conventional CPV or bifacial PV single-axis tracked systems. Assumptions: the cost of a CPV system is 50% higher than the cost of a monofacial fixed-tilt PV system, the cost of a bifacial single-axis tracked PV system is 14% higher than a monofacial fixed-tilt PV installation and the cost of hybrid CPV/PV technology is higher than the cost of a CPV system by 11% of the cost of a monofacial fixed-tilt PV installation. Note the range change in the color bar above 106%.

As shown in blue tones, hybrid CPV/PV technology is expected to deliver the lowest solar electricity generation costs in 29% of all investigated locations worldwide. Here it is up to 6% cheaper (blue pixels) in certain parts of southern Mexico, Spain, southern Africa and the Middle East. These are the areas where hybrid CPV/PV achieves the highest potential. Regions with cost advantage would extend to places with lower DNI if the cost of CPV technology can get closer to the cost of conventional PV today. At the same time, the cost advantage will shrink if PV becomes cheaper while CPV keeps the same cost. The model we present here will keep its validity and can be adapted quickly to future scenarios. Some additional details are discussed in the appendix in subsection 7.4.

## 6. Summary and Conclusion

In this article, we have presented a method to accurately model and predict the power output of hybrid CPV/PV modules. The method calculates the expected power outputs of the CPV cells and the flat plate PV part separately based on empirical models from the literature [12, 13]. These models were parametrized based on measurements performed for an EyeCon hybrid CPV/PV module in Freiburg, Germany and they were improved in this work to better account for spectral effects. In this way, their robustness is increased for their use in a worldwide calculation. Satellite meteorological data freely available from different sources, such as TMY files [36], ground albedo [39] and optimum tilt angle [36] maps were used as inputs to calculate the energy yield of the EyeCon hybrid CPV/PV technology. Our results show that these modules could annually generate up to 1150 kWh/m$^2$ in high DNI locations. Under these conditions, the integrated flat plate c-Si cells contribute in average 15.6% to the annual energy. Moreover, we show that neglecting spectral variation in our calculations results in an average energy yield overestimation of 6.2%$_{rel}$ for the 4J CPV string and 2.1%$_{rel}$ for the flat plate PV cell array. Hence, it is important to consider the effect of spectral variation in all PV technologies, but especially those that use multi-junction solar cells such as CPV and hybrid CPV/PV.

A comparison of energy yield was performed for a 4J EyeCon hybrid CPV/PV module, a 4J FLATCON® CPV module [19] and a monofacial and a bifacial c-Si PV module in a fixed-tilt orientation and with single-axis tracking. Our results indicate that hybrid CPV/PV delivers the highest annual energy yield per module area in 99.9% of the calculated locations. In fact, hybrid CPV/PV technology is expected to generate up to 34% more energy per m$^2$ than its closest contender in locations where the annual DNI and DHI/GHI ratio are around 1203 kWh/m$^2$ and 0.49, respectively. Below DHI/GHI of 0.49 the energy surplus decreases with respect to CPV, whereas above 0.49 it decreases relative to bifacial PV single-axis tracked modules. The additional energy yield predicted for hybrid CPV/PV modules can be used to understand where this technology can offer economic advantages in terms of cost per kWh. Based on a set of system cost assumptions between PV technologies we find that the largest savings in the cost of electricity can be reached in regions with medium to high DNI in southern Mexico, Spain, southern Africa and the Middle East. Here hybrid CPV/PV technology can outperform both, bifacial PV installations with single-axis tracking and conventional CPV system with dual-axis tracking. This is due to the additional use of the diffuse and rear side irradiance. Moreover, hybrid CPV/PV modules nearly always generate the highest power output per unit area. This represents an advantage as less materials intensity leads to better sustainability. Furthermore, a hybrid CPV/PV system delivers a more stable supply of electricity because during overcast or extremely windy conditions the Si PV cells continue to generate power from diffuse sunlight and prevent the inverter from shutting down.

# 7. Appendix

## *7.1 Worldwide Calculation of Spectral Parameters Z and Mismatch Factor SMM*

Solar spectral irradiance data is crucial to accurately model PV power output, however it is typically not available, if at all, with high spatial resolution. Instead, we derived and used spectral parameters $Z_{1-2}$ and $Z_{1-3}$ [14, 17, 18] and the spectral mismatch factor SMM [15] to account for the shape of the solar spectrum. Although the use of spectral matching ratios (SMR) [52] is stipulated in IEC 62670-3 [53] to quantify the spectral dependence of multi-junction CPV cells, these are correlated to the spectral parameters Z according to $Z = 1 - 2 / (1 + SMR)$. The reason we used Z parameters instead of SMR values in this work is because the applied CPV power output model [12] is based on the former. Therefore, we calculated the spectral parameters Z and the SMM factor from measurements performed in Freiburg using calibrated III-V component cells under a collimating tube and a c-Si reference cell mounted on a dual-axis solar tracker. **Equations 7** and **8** give the expressions for calculating $Z_{1-2}$ and $Z_{1-3}$, where $I_1$, $I_2$ and $I_3$ correspond to the short-circuit currents of the component cell sensors made of GaInP (bandgap $E_g$ = 1.9 eV), GaInAs ($E_g$ = 1.4 eV) and Ge ($E_g$ = 0.7 eV) under an arbitrary or the reference AM1.5d spectrum, i.e. $I_{1\_ref}$, $I_{2\_ref}$ and $I_{3\_ref}$.

$$Z_{1-2} = 2 \, \frac{I_1}{I_{1\_ref}} \bigg/ \left( \frac{I_1}{I_{1\_ref}} + \frac{I_2}{I_{2\_ref}} \right) - 1 \qquad (7)$$

$$Z_{1-3} = 2 \, \frac{I_1}{I_{1\_ref}} \bigg/ \left( \frac{I_1}{I_{1\_ref}} + \frac{I_3}{I_{3\_ref}} \right) - 1 \qquad (8)$$

In this way, $Z_{1-2}$ mainly quantifies the expected current imbalance of a 4J solar cell due to Rayleigh and Mie scattering in the range from 300 to 600 nm. Similarly, $Z_{1-3}$ quantifies the current imbalance between the 300 to 600 nm range, due to Rayleigh scattering, relative to the 900 to 1800 nm range, due to PW absorption.

For a single-junction solar cell, it is enough to use the spectral mismatch factor [15] to assess the influence of spectral variation. Therefore, we used the SMM factor of a Si reference cell identical to the ones used in the EyeCon module as a proxy for their response during outdoor measurements. The expression used to calculate the SMM factor is given in **Equation 9**, where $I_{sj}$ and $I_{sj\_ref}$ correspond to the short-circuit currents of the Si reference cell under an arbitrary or the reference AM1.5g spectrum.

$$SMM = \frac{I_{sj}}{GNI} \Big/ \frac{I_{sj\_ref}}{1000\ W/m^2} \qquad (9)$$

Although it is known that the spectral backplane irradiance tends to shift towards longer wavelengths with respect to spectral GNI, the deviation in bifacial power output is below 0.5%$_{rel}$ when assuming the same SMM for the irradiance on both sides. As explained in [54], the reason is that the backplane irradiance is typically an order of magnitude lower than the front one.

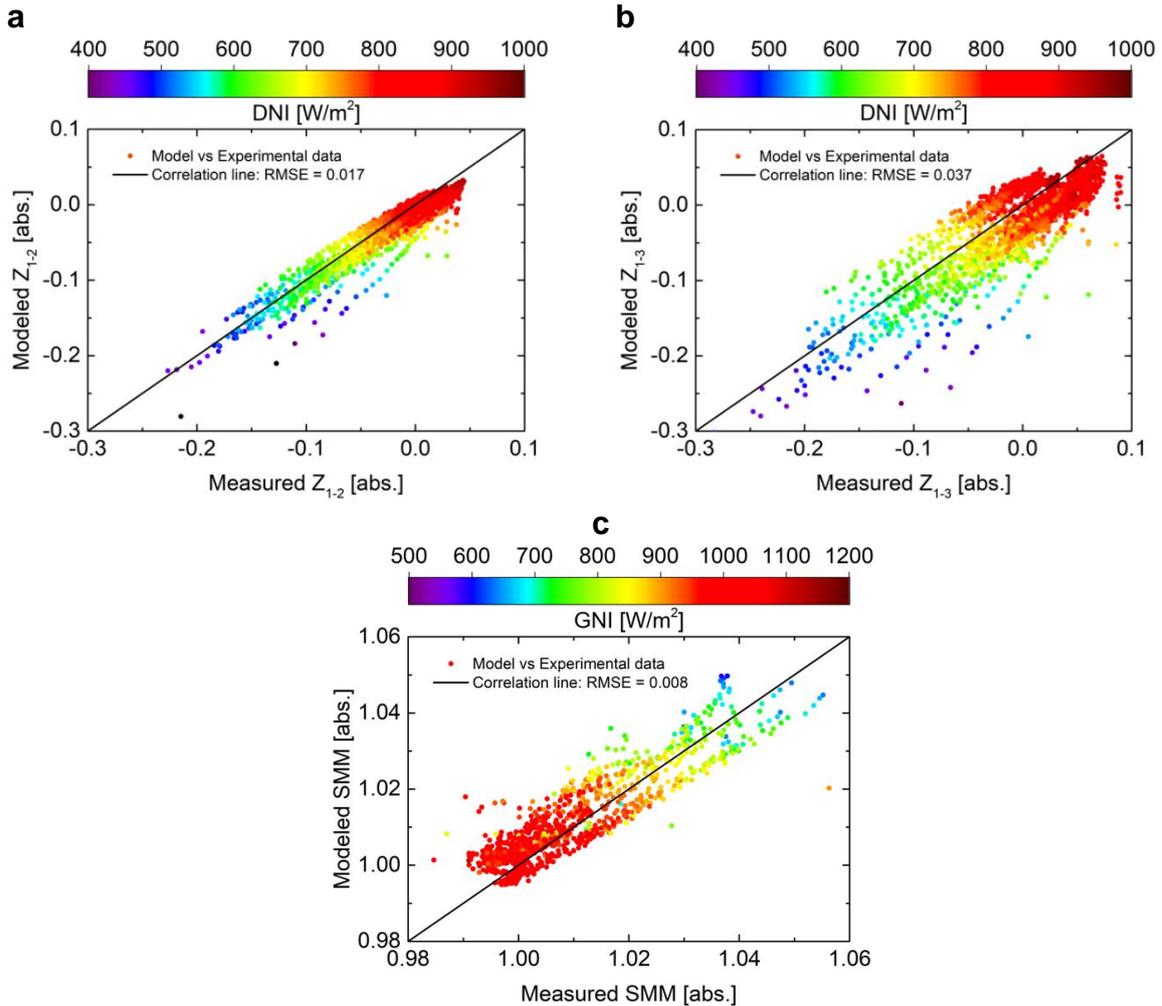

**Figure 8:** Correlation plots of modeled versus measured spectral parameters **(a)** $Z_{1-2}$ with DNI color coding **(b)** $Z_{1-3}$ with DNI color coding and **(c)** spectral mismatch factor with GNI color coding in Freiburg, Germany during the summer of 2019. The legends display the NRMSE.

In order to obtain spectral parameters for our worldwide calculation, we applied the method described in [16]. Following this approach, $Z_{1-2}$, $Z_{1-3}$ and SMM are calculated using the external quantum efficiencies (EQE) of the component and reference cell sensors in combination with the clear sky solar spectrum generated via SMARTS2 [55]. In order to cover all possible atmospheric conditions encountered worldwide, we generated over 94 000 spectra from a wide parameters-sweep that comprehends 1 < AM < 30, 0 cm < PW < 12 cm, 0 < AOD < 1 and 413 mbar < $P_{atm}$ < 1213 mbar. Based on the results, a look-up table correlates the $Z_{1-2}$, $Z_{1-3}$ and SMM as a function of AM, PW, $P_{atm}$ and DNI/DNI$_{clear\_sky}$. The latter ratio contains the influence of Rayleigh and Mie scattering, therefore hourly worldwide knowledge of AOD is not required. In this way, all the necessary parameters are available since $P_{atm}$ is given in the TMY data, PW can be calculated as a function of TMY $T_{amb}$, RH and $P_{atm}$ according to [56], AM only depends on location and solar position and DNI/DNI$_{clear\_sky}$ is computed from TMY DNI and modeled clear sky DNI using SMARTS2. DNI$_{clear\_sky}$ is only a function of AM and $P_{atm}$ since PW = AOD = 0 per definition of clear sky in this work. For the simplicity of the method its accuracy is remarkable as shown in the correlation plots of **Figure 8**. There, the estimations of $Z_{1-2}$, $Z_{1-3}$ and SMM show low RMSE of 0.017, 0.037 and 0.005 with negligible bias for a typical summer in Freiburg, Germany.

## 7.2 Multi-junction CPV Cell Modeling of $I_{SC}$/DNI

As mentioned in subsection 2.1, the measured $I_{SC}$/DNI of the 4J solar cells deviates from the modeled response for $Z_{1-2}$ values below $Z_{min}$. Using **Equation 10**, the $I_{SC}$/DNI of the 4J CPV array was modeled using the spectral response at 25ºC of each sub-cell, i.e. $SR_i(\lambda)$, and the solar spectra, $E(\lambda)$, generated from the SMARTS2 parameters-sweep in the previous subsection,

$$\frac{I_{SC}}{DNI} = min\left(\eta_{opt}\, A_{lens}\, N_p \int_0^\infty SR_i(\lambda)\, E(\lambda)\, d\lambda\right)\Big/DNI \qquad (10)$$

where $N_p$ = 4 is the number of 4J CPV solar cells interconnected in parallel and $\eta_{opt}$ = 0.85 and $A_{lens}$ = 22.7 cm$^2$ are the optical efficiency and the aperture area of the Fresnel lens, respectively. As shown in **Figure 9**, the measured $I_{SC}$/DNI versus $Z_{1-2}$ data (gray dots) agrees with the modeled values (color coded dots) even though the cell temperature was fixed at 25ºC in the model. Nevertheless, for $Z_{1-2} \leq Z_{min}$ values it is evident that the fit to the measured data (dashed line) would result in an overestimation as compared to the modeled values. Therefore, the coefficients derived from the modeled response below the minimum measured value of

$Z_{1-2}$, i.e. $Z_{min}$, are applicable for $Z_{1-2} \leq Z_{min}$. On the other hand, the coefficients derived from the measured response are applicable for $Z_{1-2} > Z_{min}$. Moreover, the apparent model discontinuity at $Z_{min}$ is due to neglection of the EQE temperature dependence. However, the temperature coefficients $i_3$ and $i_4$ derived from measured data at $Z_{1-2} > Z_{min}$ in **Equation 1** are superimposed for usage also at $Z_{1-2} \leq Z_{min}$ in order to account for temperature effects in that range.

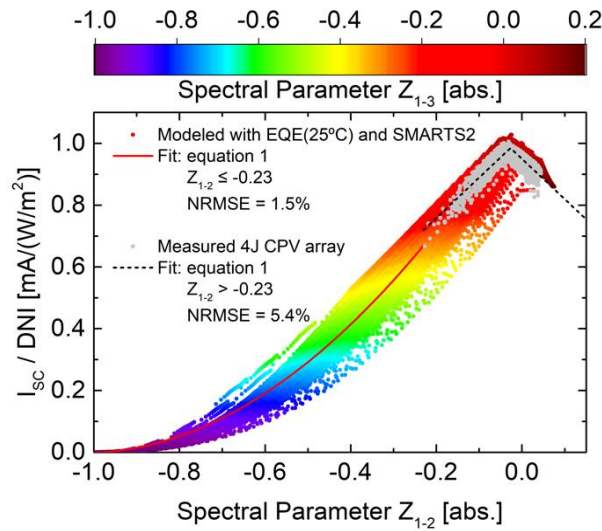

**Figure 9:** Measured (gray circles) and modeled (color coded circles according to $Z_{1-3}$) $I_{SC}$/DNI versus spectral parameter $Z_{1-2}$. The measured data agrees with the modeled values obtained with the EQE of the 4J solar cell at 25ºC and the spectral DNI generated with SMARTS2 in the range $Z_{1-2} > -0.23$. **Equation 1** was used to fit the measured (black dashed line) and modeled (solid red line) data above and below $Z_{1-2} = -0.23$ with NRMSE of 1.5 and 5.4%, respectively. The fit discontinuity at $Z_{1-2} = -0.23$ originates from neglecting the EQE temperature dependence in the model.

## *7.3 Mounting of PV Modules and Irradiance Calculation in the Plane of the Array*

For hybrid CPV/PV and standard CPV modules to reach their high performance they need to be mounted on high-precision (< 0.2º) dual-axis solar trackers. Therefore, these two technologies are always assumed normal to the sun in our calculations. In addition, the large structures (ca. 100 m$^2$) where they are mounted usually positions the modules with a significant elevation above the ground. On the contrary, flat plate c-Si PV modules are typically deployed closer to the ground on static or simpler mechanical structures with only one degree of freedom. Therefore, it is important to clarify the relevant assumptions regarding module mounting and sun tracking for hybrid CPV/PV, standard CPV and flat plate PV technologies. Regarding clearances above the ground 0, 1 and 4 m were chosen for fixed-tilt, single- and dual-axis tracking configurations, since these are typical values seen in PV installations. This corresponds to elevation to module-height ratios of 0, 0.5 and 2. In addition, a fixed-tilt PV module is assumed to face the equator at the optimum tilt angle to maximize energy yield. The world map of optimum tilt angles used here is freely available from PVGIS [36]. For a PV module with single-axis tracking, the rotation axis is aligned in the north-south direction and tilted towards the equator at the same angle as the fixed-tilt module. Moreover, axis rotation in the east-west direction is always enabled between ±90º in order to minimize the angle of incidence. In our calculation this is implemented with the single-axis tracking algorithm [57] from PV lib.

Successively, global, diffuse and backplane irradiances on a fixed-tilt, single- or dual-axis tracked plane are calculated using the algorithm in [44] that is also implemented in PV lib [58]. The method uses DNI, DHI, AM, the module orientation, the solar position [59], the anisotropic sky Perez model [60], the yearly average ground albedo map and the sky and ground view factors as inputs. Furthermore, it accounts for the reflection losses associated with the angle of incidence of direct and diffuse irradiance according to [61]. In addition, the sky and ground view factors consider sky-masking and ground-shading by the module itself [44, 62–64]. It is important to note that their calculation drastically increases the computing intensiveness due to the solution of over 1.3 billion integrals for each worldwide scenario. Nevertheless, they significantly improve the accuracy of the irradiance calculation, particularly for the backplane.

## *7.4 Calculation of Relative Cost of Hybrid CPV/PV Electricity*

In this work, the cost of electricity (CoE) of a PV technology is calculated with **Equation 11** as the ratio between the total cost of the system (CoS) and the total amount of electricity it produces throughout its lifetime.

$$CoE = \frac{CoS}{E} \quad (11)$$

In order to enable our findings to remain useful under different cost scenarios, we correlated the system cost of hybrid CPV/PV and bifacial PV single-axis tracking technologies with the cost of their simpler counterparts, i.e. conventional CPV and monofacial PV fixed-tilt systems. These cost correlations are given in **Equations 12** and **13** and according to the assumptions explained in section 5 we defined the correlation factors a and b as 1.15 and 0.11.

$$CoS_{Bifacial\ PV\ Single-Axis\ Tracking} = a \cdot CoS_{Monofacial\ PV\ Fixed-Tilt} \quad (12)$$

$$CoS_{Hybrid\ CPV/PV} = CoS_{CPV} + b \cdot CoS_{Monofacial\ PV\ Fixed-Tilt} \tag{13}$$

Furthermore, we analyze the cost of hybrid CPV/PV electricity relative to its closest competitor, i.e. $CoE_{closest}$ = min($CoE_{CPV}$ , $CoE_{Bifacial\ PV\ Single-Axis\ Tracking}$), as it is denoted by the ratio given in **Equation 14**.

$$\frac{CoE_{Hybrid\ CPV/PV}}{CoE_{closest}} = \frac{CoS_{Hybrid\ CPV/PV}}{CoS_{closest}} \cdot \frac{E_{closest}}{E_{Hybrid\ CPV/PV}} \tag{14}$$

Substitution of **Equations 12** and **13** into **14** leads to **Equation 15**. This expression shows that the cost of hybrid CPV/PV electricity relative to its closest competitor ($CoE_{rel}$) depends on the cost ratio between the reference systems, i.e. R = $CoS_{CPV}$ / $CoS_{Monofacial\ PV\ Fixed-Tilt}$, and the ratio between the electricity produced by its closest competitor and itself, i.e. $E_{closest}$ / $E_{Hybrid\ CPV/PV}$.

$$CoE_{rel} = min\left(\frac{(R+b)}{R} \cdot \frac{E_{CPV}}{E_{Hybrid\ CPV/PV}}, \frac{(R+b)}{a} \cdot \frac{E_{Bifacial\ PV\ Single-Axis\ Tracking}}{E_{Hybrid\ CPV/PV}}\right) \tag{15}$$

The systematic parametrization implemented in **Equation 15** enables the calculation of the reduction in the cost of hybrid CPV/PV electricity for different system cost scenarios. For instance, **Figure 10a** shows the mean reduction in the cost of hybrid CPV/PV electricity relative to its closest competitor as a function of R (blue line, error bars denote standard deviation), for cost scenario 1, i.e. a = 1.14 and b = 0.11. In this case, hybrid CPV/PV electricity becomes cheaper than its closest competitor when the cost of a CPV system is less than 1.7 times the cost of a monofacial PV fixed-tilt system (red line). As the R factor decreases to 1.1, the cost of hybrid CPV/PV electricity becomes up to (7.9 ± 4.8) % cheaper (blue line) in 92% of the locations investigated in the world maps (red line).

Assuming that the cost of a bifacial PV single-axis tracking system is only 9% higher than its monofacial fixed-tilt counterpart corresponds to the scenario 2, i.e. a = 1.09 and b = 0.11. As depicted in **Figure 10b**, the relative cost of hybrid CPV/PV electricity with respect to its closest competitor (blue solid line) becomes in average 1% more expensive than in scenario 1. In contrast, scenario 3 assumes that the cost of turning a conventional CPV system into a hybrid one is 16% of the cost of monofacial PV fixed-tilt technology, i.e. a = 1.14 and b = 0.16. In this case, hybrid CPV/PV electricity (blue dashed line) becomes in average 1.6% more expensive than in scenario 1. Therefore, an increase in the cost of a hybrid CPV/PV system is more influential than a proportional decrease in the cost of a bifacial PV single-axis tracking system.

In addition, for scenarios 2 and 3 the cost of hybrid CPV/PV electricity begins to become cheaper below R values of 1.6. However, the percentage of locations where this occurs (red lines) decreases in average by 11 and 19% relative to scenario 1. The value of calculating scenarios 1 to 3 consists in covering the range where the relative cost of hybrid CPV/PV electricity could fall as the energy market continues to develop.

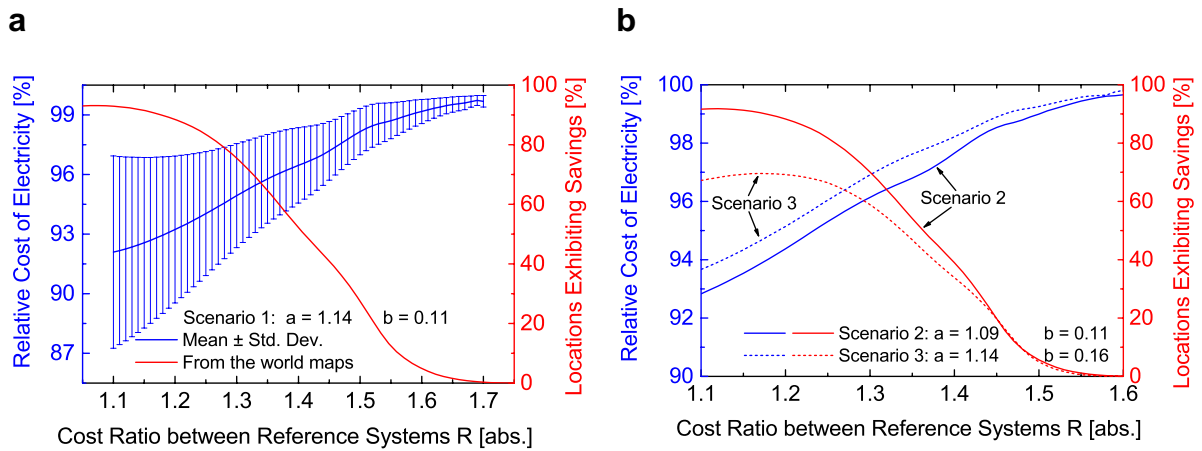

**Figure 10: (a)** Plot of the mean relative cost of hybrid CPV/PV electricity with respect to its closest competitor (blue line), i.e. whichever is cheaper between CPV and bifacial PV single-axis tracking electricity, and of the percentage of locations where hybrid CPV/PV beats them (red line), as a function of the cost ratio between the reference systems R, i.e. CPV and monofacial fixed-tilt PV technology. The plots correspond to scenario 1, where a = 1.14 and b = 0.11 and the error bars denote the standard deviation for each value of R. **(b)** Same as (a) but for scenario 2, where a = 1.09 and b = 0.11, and for scenario 3, where a = 1.14 and b = 0.16, without standard deviations to avoid clutter.

## Acknowledgement

The authors acknowledge the "III-V Photovoltaics and Concentrator Technology" department at Fraunhofer ISE for their support regarding the development of the bifacial hybrid CPV/PV EyeCon module. Additionally, the authors acknowledge the PVGIS website for making freely available all the meteorological data used in our calculations. This work has been financially supported in part by the National Council of Science and Technology (CONACYT) and the Mexican Secretary of Energy (SENER) in the form of a Ph.D. scholarship for Juan F. Martínez. The authors are responsible for the contents of this paper.